%% file: paper.tex
\documentclass[10pt,a4paper,twocolumn]{article}

\usepackage[margin=1in]{geometry}
\usepackage{times}
\usepackage{amsfonts}
\usepackage{color}
\usepackage{color}

\usepackage{graphicx}
\usepackage{caption}
\usepackage{subfig}
\usepackage{epsfig} % for postscript graphics files
\usepackage{amsmath}
\usepackage{amssymb}
\usepackage{multirow}
\usepackage[table]{xcolor}
\usepackage{url}
\usepackage[font=small,labelfont=bf]{caption}

\DeclareMathOperator*{\argmin}{arg\,min}

\title{Dynamic Group Behaviors for Interactive Crowd Simulation}

\author{Liang He$^1$ and Jia Pan$^2$ and Sahil Narang$^1$ and Wenping Wang$^2$ and Dinesh Manocha$^1$
 \thanks{$^1$ Liang He, Sahil Narang, and Dinesh Manocha are with the Department of Computer Science, the University of North Carolina at Chapel Hill}
  \thanks{$^2$ Jia Pan and Wenping Wang are with the Department of Computer Science, the University of Hong Kong}
}

\date{}

\begin{document}

\twocolumn[{%
\renewcommand\twocolumn[1][]{#1}%
\maketitle
\begin{center}
    \centering
    \includegraphics[width=0.32\linewidth]{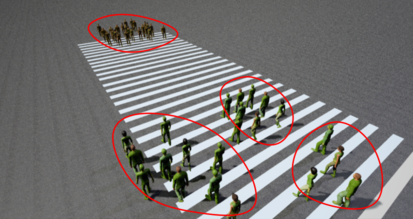}
    \includegraphics[width=0.32\linewidth]{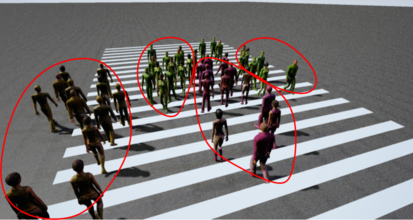}
    \includegraphics[width=0.32\linewidth]{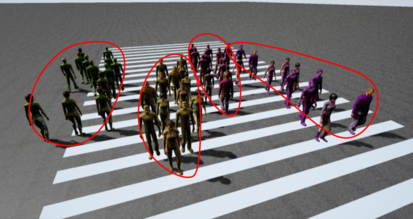}
    \captionof{figure}{Our method can generate dynamic group behaviors along with coherent and collision free navigation. We highlight the performance in a street-crossing scenario, where different groups are shown with different colors. Our approach can automatically adapt to the environment and the  number, shape, and size of the groups can change dynamically.}
   \label{fig:benchmark1}
\end{center}%
}]

\begin{abstract}
We present a new algorithm to simulate dynamic group behaviors for interactive multi-agent crowd simulation. Our approach is general and makes no assumption about the environment, shape, or size of the groups. We use the least effort principle to perform coherent group navigation and present efficient inter-group and intra-group maintenance techniques. We extend the reciprocal collision avoidance scheme to perform agent-group and group-group collision avoidance that can generate collision-free as well as coherent and trajectories. The additional overhead of dynamic group simulation is relatively small. We highlight its interactive performance on complex scenarios with hundreds of agents and compare the trajectory behaviors with real-world videos.
\end{abstract}

%\IEEEpeerreviewmaketitle

\input{intro.tex}
\input{related.tex}

\input{overview.tex}

\input{method.tex}

\input{experiment.tex}

\input{conclusion.tex}

\bibliographystyle{IEEEtran}
\bibliography{references}

\end{document}

%% file: intro.tex
\section{Introduction}
\label{sec:intro}

The problem of simulating the trajectories and behaviors of a large number of human-like agents frequently arises in computer graphics, virtual reality, and computer-aided design. It includes generation of pedestrian movements in a shared space and the collaboration between the agents governed by  social norms and interactions. The resulting crowd simulation algorithms are used to generate plausible simulations for games and animation, as well as  accurately predicting the crowd flow and patterns in architectural models and urban environments.

One of the main challenges is modeling different behaviors corresponding to navigation, collision-avoidance, and social interactions that lead to  self-organization and emergent phenomena in crowds.  
Prior research and observations in sociology and behavioral psychology have suggested that real-world crowds are composed of (social) groups.  The group is a meso-level concept and is composed of two or more agents that share similar goals, over a short or long period of time, and exhibit similar movements or behaviors. In many instances, up to 70\% of observed pedestrians are walking in such groups~\protect\cite{coleman1961equilibrium,Gorrini:2014:dynamics}. As a result, it is important to model the group dynamics that includes intra-group and inter-group interactions within a crowd.

In this paper, we address the problem of simulating the group behaviors that are similar to those observed in real-world scenarios. In  crowds, small groups are dynamically formed  as some of the agents move towards their goals and generate behaviors such as aggregation, dispersion, following the leader, etc. As the individual agents  respond to a situation (e.g. panic or evacuation), the dynamic behaviors can result in splitting a large group or new groups being formed.
Such group behaviors are frequently observed in public places, sporting events, street-crossing, cluttered areas when the pedestrians tend to avoid the obstacles, etc. 
Furthermore, the geometric shape of the group and the size of these groups may change. Some earlier observations have suggested that group sizes different according to a Poisson distribution~\protect\cite{james1953distribution}.

Prior work on modeling group behaviors is mostly limited to cohesive movements or spatial group structures. The simplest algorithms cluster the agents into a fixed number of groups and the size of each group remains fixed (i.e. static grouping). They are unable to model the changing shape or size of the group, splitting of a large group into sub-groups or merging of smaller groups into a large group. Furthermore, in some scenarios an agent may switch from one group to the other group in close proximity. It is important to model such dynamic behavior in arbitrary environments.

\noindent{\bf Main Results:} We present a novel algorithm to generate dynamic group behaviors using multi-agent crowd simulation. Our approach is general and makes no assumptions about the number, size, or shape of the groups. We use spatial clustering techniques to generate group assignments that take into account the positions and velocities of the agent. Our group-level navigation algorithm is based on the principle of least effort that tends to maintain the group relationships as each agent proceeds towards its goal position. We present efficient inter-group and intra-group level techniques to perform coherent and collision-free navigation. The group shape and sizes are automatically updated as new agents are assigned to the group or when some agents leave the group.

We extend the agent-agent reciprocal collision avoidance algorithm~\protect\cite{Jur:2011:RVO} to perform agent-group and group-group reciprocal collision avoidance. We formulate the velocity obstacle of the group in terms of the convex hull of the current agent positions and use that to perform conservative collision avoidance. Our approach is used to compute the new preferred velocity for each agent that not only avoids collisions with other agents and obstacles, but also performs coherent group navigation. This makes it possible to handle high-density crowds as well as arbitrarily shaped groups.

The overall approach has been implemented and we highlight its performance on many complex benchmarks with dynamic group behaviors. The additional overhead of group computation and maintenance is relatively small and our approach takes a few milli-seconds per frame on scenarios with hundreds of agents. Our formulation can generate smooth and coherent group-level trajectories and we demonstrate the benefits over prior methods based on agent-based  or meso-scale algorithms. We compare the trajectory behaviors generated by our algorithm with real-world crowd videos by extracting the pedestrian trajectories. 
Overall, our approach offers the following benefits:

\begin{enumerate}
\item Our approach is general and makes no assumption about the environment, size or shape of the groups.
\vspace*{-0.1in}
\item We present an efficient algorithm for agent-group and group-group collision avoidance by extending the reciprocal velocity obstacle formulation.
\vspace*{-0.1in}
\item Our approach can generate dynamic group behaviors in terms of formation, merging, splitting, and re-assignment.
\vspace*{-0.1in}
\item We observe plausible group behaviors and smooth trajectories, similar to those observed in real-world crowds.
\end{enumerate}

The rest of the paper is organized as follows. We briefly survey prior work in crowd simulation and group behaviors in Section 2. We introduce the notation and give an overview of our approach in Section 3. The overall algorithm is described in Section 4, and we highlight its performance in Section 5.

%% file: related.tex
\section{Related Work}
\label{sec:related}
In this section, we give a brief overview of prior work on crowd simulation and group behaviors.

\subsection{Crowd Simulation}
There is extensive work on modeling the behavior of crowds. These include multi-agent simulation techniques for computing collision-free trajectories and navigation based on   social forces~\cite{helbing1995social}, rule-based methods~\cite{Reynolds1987,HiDac}, geometric optimization~\cite{Jur:2011:RVO,Pettre:2009:EMS,Karamouzas:2012:SEL}, vision-based steering~\cite{Ondrej:2010:SBS}, cognitive methods~\cite{YuTerzopoulos}, personality models~\cite{UPennOCEAN}, etc. Other class of simulation algorithms are based on data-driven methods~\cite{Lerner:2009:DDE,KapadiaDatabase}  and estimating the simulation parameters based on real-world crowd data~\cite{Wolinski2014,KapadiaSteerFit}. The macroscopic  simulation algorithms compute fields for pedestrians to follow based on continuum flows~\cite{TreuilleContinuum} or fluid models~\cite{RahulCrowd} and are mainly used for high-density crowds.

\subsection{Group Behavior Simulation}
Group behaviors have been studied in computer graphics~\cite{Lee07,park2012modeling,Curtis:2012:WPE,Huerre:2010:SBC:1900520.1900533}, robotics~\cite{Krontiris:2012:ICRA}, 
pedestrian dynamics~\cite{Gorrini:2014:dynamics}, and social psychology~\cite{knowles1973boundaries}. Prior techniques have been mainly used to simulate  static or fixed-sized  groups and perform group-based collision avoidance~\cite{Santos:2014:Robotica,He:2013:ICRA,Karamouzas:2015:ICRA}. 
However, none of these methods can efficiently simulate dynamic groups of varying sizes in arbitrary environments. Golas et al.~\cite{golas2014hybrid} have proposed a hybrid approach that combines microscopic and macroscopic methods, and generates grouping behaviors by taking into account  long range trajectory predictions. 
%This is based on the fact that nearby agents tend to have similar future predictions, which will push nearby agents together during the navigation. 
However, this approach cannot generate stable grouping behavior, and long range prediction can be expensive. 
Recently, a distributed following strategy~\cite{He:2015:PGB} has been proposed for dynamic behaviors, but is limited to scenes with a few agents and cannot simulate arbitrary merging and splitting behaviors or handle large number of groups.

%% file: overview.tex
\section{Overview}
\label{sec:overview}

In this section, we introduce our notation and give an overview of our approach.

\subsection{Dynamic Grouping Behavior}

\begin{figure}[t]
  \centering
  \includegraphics[width=\linewidth]{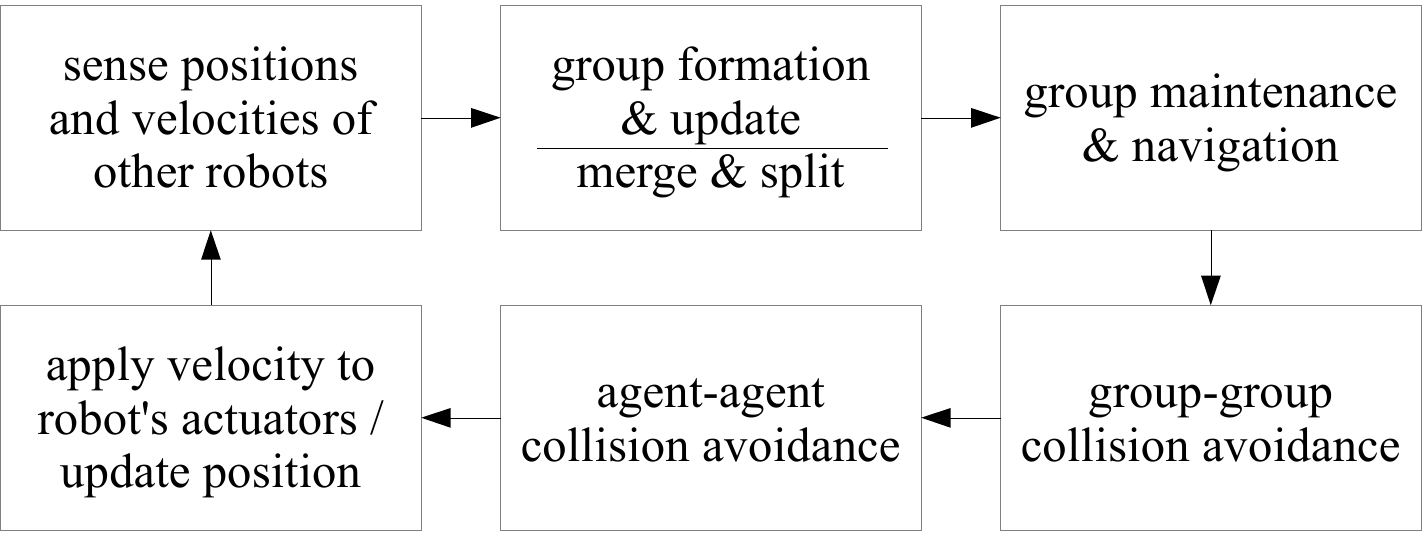}
  \caption{Algorithm pipeline: We show the various components of our algorithm for dynamic group behaviors.}
  \label{fig:pipeline}
\end{figure}

Our approach is designed for multi-agent crowd simulation algorithms. We assume that during each step of the simulation, each agent in the crowd has an intermediate goal position that is used to compute its preferred velocity. This goal position can change over the course of the simulation. 
The notion of dynamic grouping is motivated by real-world crowd observations. Many  studies have highlighted the importance of group dynamics in the context of modeling the interactions between the agents and with the objects in the environment~~\protect\cite{reicher2001psychology}.  The number of such groups or the size of each group (i.e. number of agents) can change during the course of the simulation. 

The dynamic grouping behavior within a crowd is classified based on how the agents are dynamically clustered into groups. 
Given a set of $n$ independent agents sharing a (2D) environment consisting of obstacles, we automatically compute these groups using spatial and temporal clustering algorithms. In particular, given the current position $\mathbf p_a$ and velocity $\mathbf v_a$ of an agent $a$, we need to cluster all agents into a set of groups $\{G^i\}$ according to a pairwise similarity metric defined over the agents. It is possible that some agent may not belong to any group and is treated as an isolated agent. The specific group assigned to an agent $a$ is denoted as $G_a \in \{G^i\}$. We also compute the velocity of a group $G$ as the average velocity of all agents belonging to $G$, and is denoted as $\mathbf v_G$. During the simulation, the number of agents in a group $G$ may change or or may maintain the group formation. For example, nearby agents with similar goals and similar directions of motion will merge into a group and maintain the group by following one after another.
A large group may split into several sub-groups while facing an obstacle or other groups, and these sub-groups may merge into one group after passing the obstacle. As two groups come close to each other, it is possible that agents may switch from one group to the other (i.e. reassign groups for an agent). As a result, it is important to support such group behaviors corresponding to formation, merging, splitting, reassignment, etc. 

\subsection{Agent-Group Velocity Obstacle}
\begin{figure}[!ht]
\centering
\includegraphics[width=0.8\linewidth]{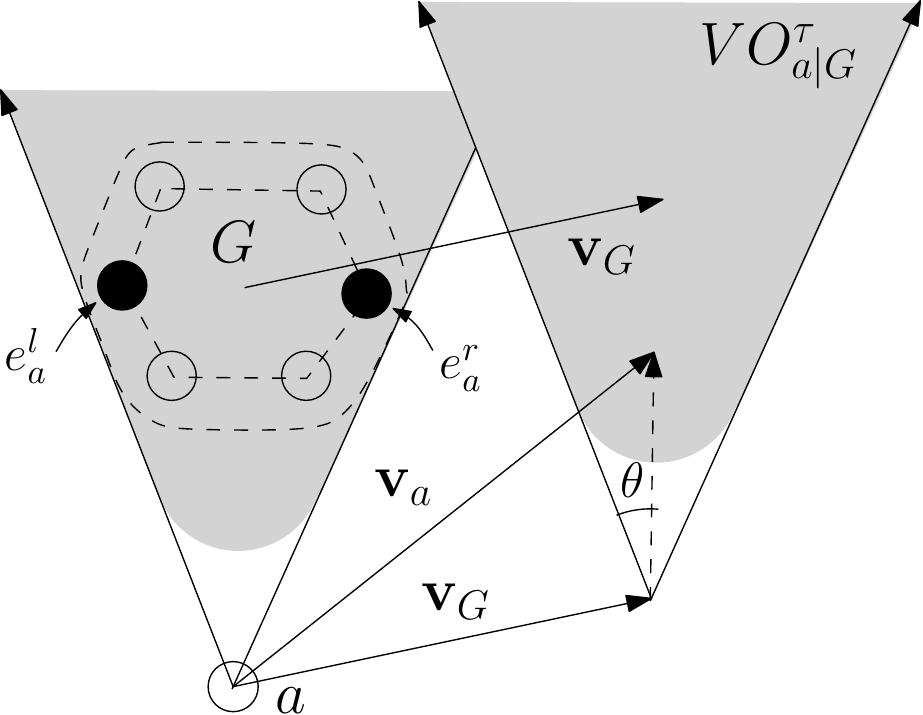}
\caption{{\bf Agent-group Velocity Obstacle:} The velocity obstacle $VO_{a|G}^{\tau}$ for agent $a$ induced by a group $G$ of agents. The group $G$ contains six agents and its convex hull is the dashed line. If $G$ only contains a single agent, $VO_{a|G}$ reduces to the traditional velocity obstacle between two agents. The black agents $e_a^l$ and $e_a^r$ are two most extreme agents in the group $G$. The angle $\theta$ represents the steering angle required by agent $a$ to avoid the group of agents $G$.}
\label{fig:rvo}
\end{figure}

For collision avoidance between the agents, we use the concept of velocity obstacles~\protect\cite{Jur:2011:RVO}. In order to perform collision avoidance between groups, we use the notion of velocity obstacle $VO_{a|G}$ for one agent $a$ induced by a group $G$. Given the velocity of the group $\mathbf v_G$, $VO_{a|G}$ can be defined as the set of agent $a$'s velocities $\mathbf v_a$ that will result in a collision with $G$ at some point within time window $\tau$ assuming that the group $G$ maintains its velocity $\mathbf v_G$ during $\tau$:
\begin{align}
\label{eq:VOaG}
VO_{a|G}^{\tau} = \{\mathbf v | & \exists t\in[0, \tau] \text{ such that } \notag \\
& \mathbf p_a + (\mathbf v - \mathbf v_G)t \in \mathcal{CH}(G) \oplus \mathcal{D}(\mathbf 0, r_a)\},
\end{align}
where $\mathcal{D}(\mathbf 0, r_a)$ is a disc centered at the origin with radius $r_a$, and $\mathcal{CH}(G)$ is convex hull of the set of agents constituting the group $G$. The convex hull provides a conservative bound that can guarantee collision free navigation.
This equation implies that if agent $a$ chooses a velocity outside the velocity obstacle $VO_{a|G}$, it will not collide with group $G$ within the time window $\tau$. Intuitively, the velocity obstacle can be geometrically constructed as a cone with apex in $\mathbf p_a$ and its sides tangent to $\mathcal{CH}(G)$ expanded by the radius $r_a$ of the agent $a$, which is then translated by $\mathbf v_G$, as shown in Figure~\ref{fig:rvo}. From the geometric interpretation, we can observe that the convex hull $\mathcal{CH}(G)$ need not be computed explicitly, instead the velocity obstacle can be fully defined by the extreme agents in the  radial directions of the group, as observed from $\mathbf p_a$. We denote the most "clockwise" agent as $e^r_a$ and the most "counterclockwise" agent as $e^l_a$. These two agents $e^r_a$ and $e^l_a$ are used to compute collision free trajectory of   agent $a$. 

\subsection{Our Approach}
Our goal is to generate realistic dynamic grouping behaviors for pedestrians.  We assign the agents to different group during each frame  and compute its trajectories by taking into account group dynamics. In cluttered areas, the agents tend to be assigned to a large group, and in open areas the agents tend to be well spread out and may not be assigned to any group. As a result, we need the capabilities to support such dynamic merging and splitting behaviors.

Furthermore, our approach is motivated by the principle of least effort~\protect\cite{guy2010pledestrians} that has been used for computing the agent trajectories in prior crowd simulation algorithms. An agent aligns itself with a group such that the resulting motion is governed by effort minimization. In particular, given the preferred velocity for each agent $a$, we tend to compute the actual velocity that tends to minimize the effort required by the entire group $G_a$ to avoid the  obstacles. In order to support dynamic groups, our approach supports the following computations:  

\noindent{\bf Group formation}: We use spatial clustering algorithm to generate the initial group assignment for each agent in the crowd. The isolated agents are not assigned to any group. 

\noindent{\bf Group maintenance and navigation}: Our approach tends to  maintain these groups as long as possible during the navigation. All the agents belonging to a group exhibit coherent trajectories and behaviors. We perform inter-group and intra-group computations to generate such behaviors. At the inter-group level, each group needs to perform high-level coherent trajectory computation  to avoid collisions with other groups and obstacles. The collision avoidance policy is chosen in a manner that if all agents in the same group consistently make the same choice, the entire group tends to avoid other groups altogether. At the intra-group level, each agent inside a group (except the group leader) will choose one fellow agent from the same group, and follow it  to make progress towards the goal. If each agent in the group follows this policy, our approach doesn't need to explicitly check for agent-agent collisions within a group.    

\noindent{\bf Group update}: The group assignments are updated and the number of agents belonging to a group may change.

A key component for trajectory computation is an efficient group-group collision avoidance algorithm. In our approach, this is achieved by first avoiding the collisions between the group leader of each group and other groups, and then determining a suitable preferred velocity for other non-leader agents. In particular, we use OCRA-based agent-group collision avoidance technique~\protect\cite{Jur:2011:RVO} to compute the velocity for the group leader. All the other agents in the same group will compute their velocity according to the following policy. 
The new adapted preferred velocity for each agent is  used by the agent-agent OCRA algorithm to compute the actual velocity for each agent by taking into account all the constraints. The preferred velocity is chosen such that it guides the agent towards its goal position.  The various components of  our approach are shown in Figure~\protect\ref{fig:pipeline}.

%% file: method.tex
\section{Multi-agent Simulation}
\label{sec:method}

In this section, we present our multi-agent simulation algorithm that can simulate dynamic grouping behaviors.   

\begin{figure}[!ht]
\centering
\subfloat[group formation]{\includegraphics[width=\linewidth]{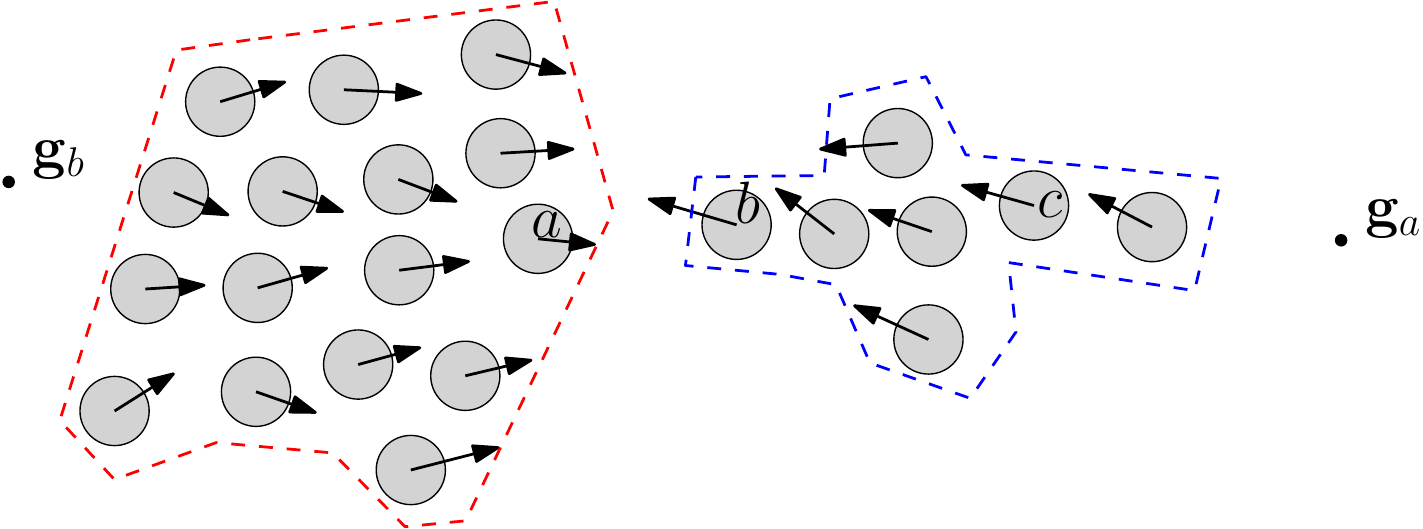}} \\
\subfloat[group maintenance]{\includegraphics[width=\linewidth]{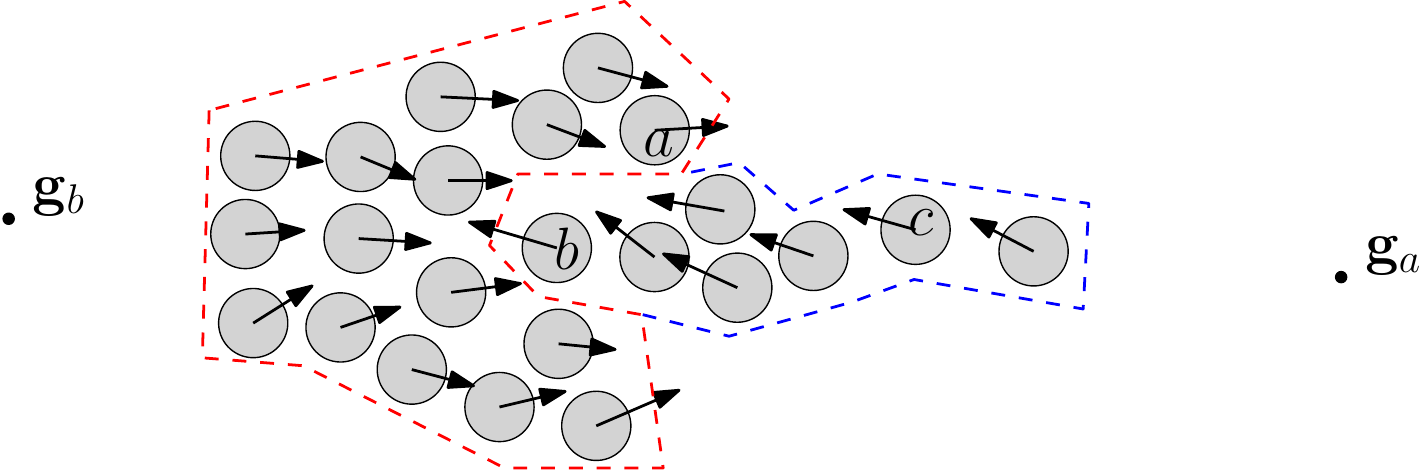}} \\
\subfloat[group split]{\includegraphics[width=\linewidth]{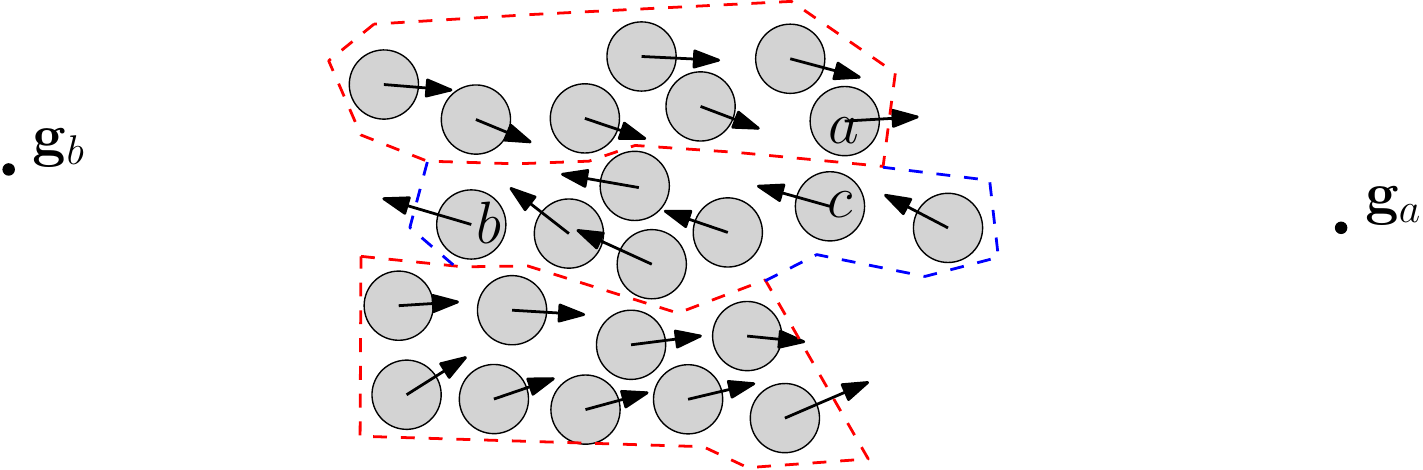}} \\
\subfloat[group re-merge]{\includegraphics[width=\linewidth]{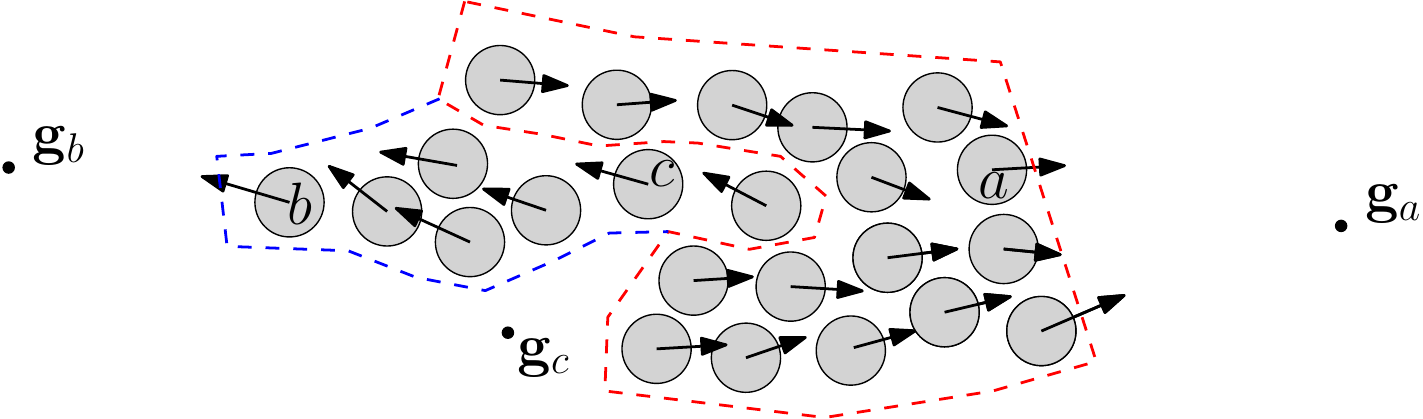}} \\
\subfloat[leave a group]{\includegraphics[width=\linewidth]{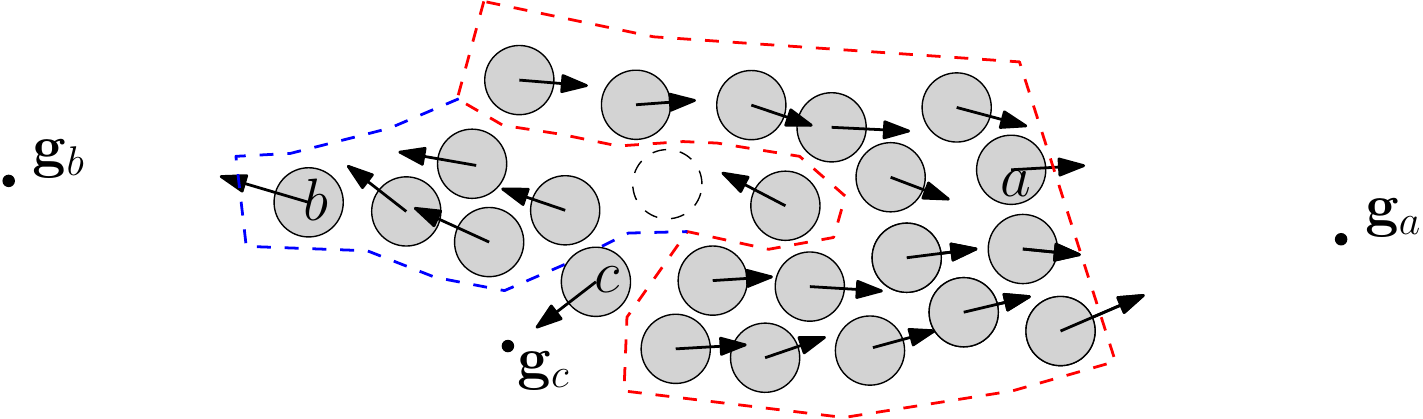}}
\caption{Dynamic group behaviors during the navigation. (a) The agents first are clustered into groups according to their position and velocities. Each agent will has its individual goal, e.g., agent $a$，$b$, $c$'s goals are $\mathbf g_a$, $\mathbf g_b$, $\mathbf g_c$ (please see (e)), respectively. (b) After a while, two groups will run into each other, but both groups will maintain their constitution during the navigation. (c) For collision avoidance, one group (marked by the red dashed line) is split into two groups. (d) After these two groups pass through each other, the split groups merge back into a single group. (e) If one agent in a group can approach its goal easily, it will choose to leave the group and navigate alone.}
\label{fig:groupbehavior}
\end{figure}

\subsection{Group Formation}
We use spatial clustering algorithm to compute the initial group assignment for each agent. This assignment is based on the positions and velocities of all agents. The clustering criteria is based on the following criteria. Given a pair of agents, $a$ and $b$, they belong to the same group if the following conditions hold:
\begin{itemize}
\item the position $\mathbf p_a$ of agent $a$ and the position $\mathbf p_b$ of agent $b$ are within a predefined distance $\epsilon_p$, and 
\item the velocity $\mathbf v_a$ of agent $a$ and the velocity $\mathbf v_b$ of agent $b$ are within a predefined threshold $\epsilon_v$.
\end{itemize}
The transitive closure of this relation uniquely classified each cluster into groups, and can be formally described as \begin{equation}
(a \sim b) \equiv (\|\mathbf p_b - \mathbf p_a\| < \epsilon_p \wedge \|\mathbf v_b - \mathbf v_a\| < \epsilon_v), \notag
\end{equation}
where $\sim$ is the binary operator defining whether two agents would be grouped together. Given this criteria for grouping, we use a greedy algorithm to compute these groups $\{G^i\}$ in $\mathcal O(nk)$ time, where $n$ are the number of agents in the crowd and $k$ is the number of groups in the partition. In particular, we iteratively check each agent whether it can be grouped into any existing groups according to the $\sim$ relationship. If an agent is not assigned to any group, it is treat as a single or isolated agent during that frame.

\subsection{Group Maintenance and Navigation}
One key point in simulating the group behavior for a crowd is how to maintain the groups based on collision avoidance constraints during the navigation. We achieve the group maintenance by using a two-level approach: the inter-group level makes sure that the entire group will avoid other groups as a whole, and the intra-group level ensures that all the agents belonging to a group do not collide with each other.

\subsubsection{Inter-Group Level}
In most multi-agent simulation algorithms,
each agent independently computes its current velocity for collision avoidance. However, such navigation algorithms may not be able to maintain the group-like coherent motion. This is because each agent may choose different extreme agents (as shown in Figure~\ref{fig:rvo}) from the same group to avoid collision, due to their difference in positions and velocities relative to the obstacle group. Instead, we would like that each agent in the same group as $a$ should select the identical side (all $e_l$ or $e_r$) while bypassing one group $G$. 

For this purpose, we first estimate the effort required for agent $a$ to bypass one obstacle group $G$ as 
\begin{align}
E_a = (\mathbf v_a - \mathbf v_G) \times (\mathbf p_a - \mathbf p_G) \cdot \mathbf n,
\end{align}
where $\mathbf v_G$ and $\mathbf p_G$ are the average velocity and position of the group $G$ respectively, and $\mathbf n$ is the normal of the 2D plane. As shown in Figure~\ref{fig:rvo}, this effort measurement is the sine function with the steering angle $\theta$ required by the agent to avoid with the obstacle group. Then the total effort for the entire group $G_a$ can be computed as $E = \sum_{b \in G_a} E_b$, and the bypassing side (for navigation) is computed as:
\begin{align}
\label{eq:side}
s =
  \begin{cases}
    \text{r (right)}   & \quad \text{if } E <0 \\
    \text{l (left)}  & \quad  \text{otherwise}.
  \end{cases}
\end{align}
In other words,  each agent would choose the same bypassing side which has a smaller effort for collision avoidance. The solution of Equation~\ref{eq:side} provides an initial direction of motion for each agent. In this way, the group $G_a$ will avoid the group $G$ as a whole.

\begin{figure}[!ht]
  \centering
  \includegraphics[width=0.7\linewidth]{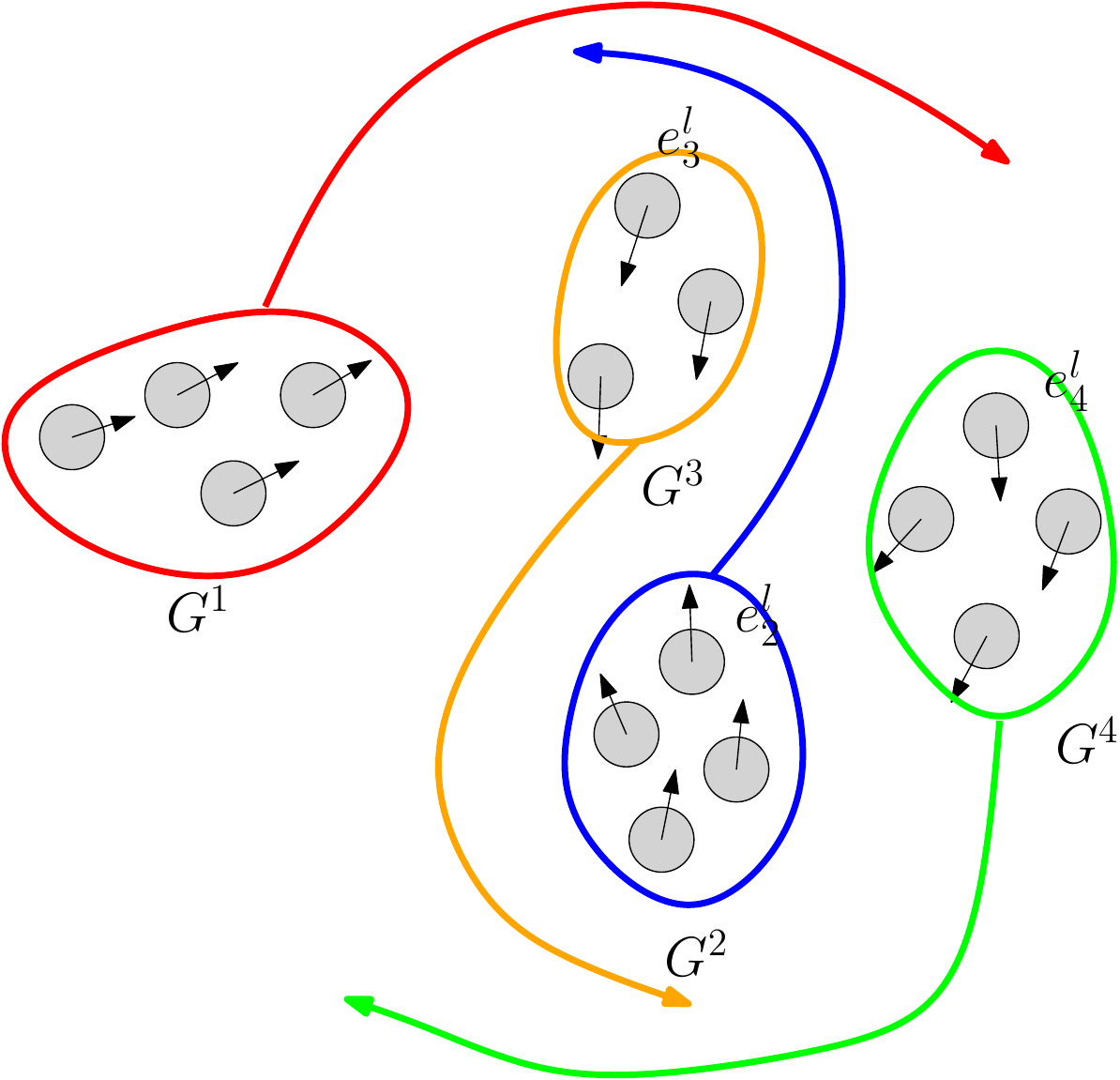}
  \caption{{\bf Group-group collision avoidance}: Our approach can compute collision-free as well as coherent trajectories for agents in each group.}
  \label{fig:multigroup}
\end{figure}

When the group $G_a$ needs to avoid a set of different groups $\{G^i\}$, it first randomly selects one group $G^i$ that may collide with it, and then computes the bypassing side $s$ and the extreme agent $e^s_i$ for it. Next, it repeatedly checks whether there are any other groups that may collide with it on the side $s$. If yes and suppose that particular group is $G^j$, then it will choose to bypass $G^j$ also from the side $s$ and the corresponding extreme agent is $e^s_j$. If not, then the iteration stops and the  extreme agent is computed. One example for this process is illustrated in Figure~\ref{fig:multigroup}. Suppose we are computing the bypassing side and extreme agent for group $G^1$ which first chooses to avoid $G^2$. It decides to bypass from the left side and the corresponding extreme agent $e^l_2$. Since both $G^4$ and $G^3$ are both to the left side of $e^l_2$, we need to further check for collision avoidance. Lets assume that we select group $G^3$ during the next step. To be consistent with the decision of avoiding $G^2$, we continue to bypass group $G^3$ from the left side, and choose $e^l_3$ as the extreme agent. Since there is no more groups to the left of $e^l_3$, the iterative process stops. In this way, group $G^1$ will bypass group $G^3$ from the left side, as shown by the red trajectory in Figure~\ref{fig:multigroup}. The trajectories for other groups can be computed in a similar manner.

\subsubsection{Intra-Group Level}
\label{sec:intra}
After computing the bypassing side for the entire group, we can achieve coherent navigation within a group. However, this may not be sufficient to avoid the reassignment, i.e., the exchange of agents between different groups.
First, the bypassing decision in the inter-group level depends on the extreme agents of a group, which is computed based on the group's convex hull (see Equation~\ref{eq:VOaG} and Figure~\ref{fig:rvo}). If the convex hulls of different groups overlap with each other, some agents in the same group may be isolated by the agents from other groups. In some other cases, the group needs to deform its shape (e.g., from a circle shape into a line shape) in order to maintain the group coherence for navigation a cluttered environment, and thus bypassing other groups from the same side may not be sufficient.

To reliably avoid group reassignment, we need to keep agents connected during the navigation. In order to simulate this trajectory behavior, we use the dynamic following strategy. In particular, we let each agent dynamically follow some other agents in the same group whenever possible. In this way, the members in a group will move along the same local path and will have the minimal risk for group reassignment or collisions with other agents. To achieve this behavior, we first need to decide whether one agent should be a leader or a follower in the group, and if it is a follower, we need to determine whom it should follow. Suppose the group $G$ chooses to bypass another group from the side $s$, then $G$'s member agents all have $e^s$ -- the extreme agent in the obstacle group on the side $s$ -- as the temporary goal $\mathbf g_G$. We choose the leader of group $G$ as the member that  is closest to $e^s$, i.e., $\text{leader} = \argmin_{a \in G} \|\mathbf p_a - \mathbf g_G\|$. All other members would be treated as the followers. 

If an agent $a$ is a follower, we choose its following target as follows. First, we find all agents $b$ in the group that satisfy $\|\mathbf p_b - \mathbf g_G\| < \|\mathbf p_a - \mathbf g_G\|$, and the set of all qualified agents is denoted as $F$. In order to compute a stable connected group, we choose $a$'s following target as the one in $F$ that is closest to $a$. This is because if $b$ is too far away from $a$ then when tried to follow $b$, the group shape may change considerably, which makes it difficult to perform group-level collision avoidance. Formally, $a$'s following target is selected as $b^* = \argmin_{b\in F, b\neq a} \|\mathbf p_b - \mathbf p_a\|$.

\subsection{Group Update}
The group update or reassignment happens under two situations. 
The first situation is while the agents are in the open area and can easily approach their goals. In this case, the group bypassing and dynamic following strategies are usually sub-optimal for an individual agent's trajectory, even though they are beneficial for the overall navigation. As a result, the notion of being able to stop following at the suitable time will help improve the performance of multi-agent navigation system. We perform this step by checking whether the original preferred velocity $\mathbf v_{\text{pref}}$ will result in making the agent collide with any other agents. If not, the agent will detach from the group and uses the discrete agent local navigation algorithm based on ORCA to move towards its goal.

The second situation arises when the current group setting is not able to compute a collision-free velocity for the navigation. This is mainly because the original groups have deformed too much during the navigation, and their shapes are become quite non-convex. Our solution is to perform re-clustering over the entire crowd, to generate a group partition that can better describe the current dynamic behavior of the pedestrian crowd. 

\subsection{Collision Avoidance}
Besides the high-level grouping behaviors, we also need to make sure that there is no collision between the individual agents in the crowd. However, prior  agent-agent collision avoidance schemes such as ORCA~\protect\cite{Jur:2011:RVO} or social forces~\protect\cite{helbing1995social} may not maintain the group assignment. Some recent methods~\protect\cite{Santos:2014:Robotica} extend the traditional agent-agent velocity obstacle by considering each group as a super-agent. However, they assume the group shape and size is should be fixed during the navigation, and thus requires all agents in the group must always choose the same velocity. As a result,  simple extensions of velocity obstacle may not be able to find a feasible solution and does not work in cluttered environments where group deformation and/or reassignment are necessary for collision avoidance. 
Instead, we use a two-level to keep grouping behavior and make sure safe navigation simultaneously.

\subsubsection{Group-Group Collision Avoidance}
For the group-group collision avoidance, we leverage the result from the following strategy in Section~\ref{sec:intra}. Given the leader agent $a$ of one group $G_a$, we first compute the adapted preferred velocity of $a$ that can avoid all the other groups. 

\begin{figure}[!!ht]
  \centering
  \includegraphics[width=0.8\linewidth]{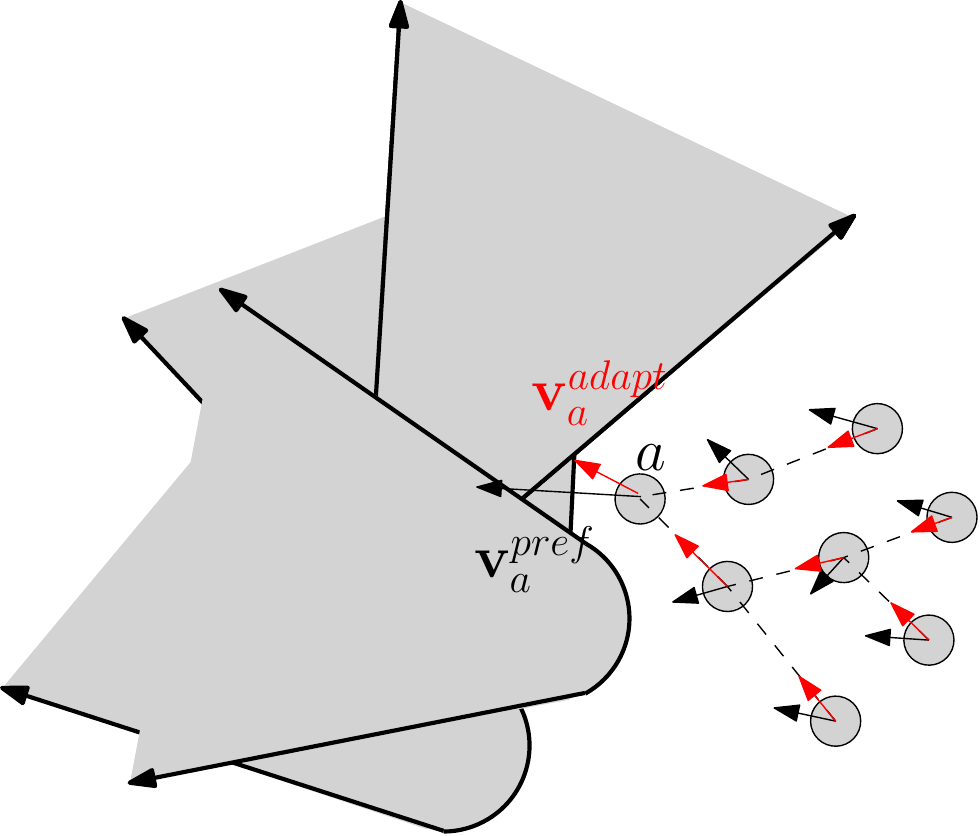}
  \caption{Group-group collision avoidance:  $a$ is the leader and the dashed lines illustrate the following relationship. The black vectors are the input preferred velocities, and the red vectors are the new preferred velocities  computed by our algorithm. They tend to avoid collisions with the other agents and used for coherent group navigation.}
  \label{fig:meso}
\end{figure}

In order to avoid the collision with other groups within time $\tau$, the agent $a$ should choose the actual adapted velocity $\mathbf v_a^{\text{adapt}}$ that is outside the union of the velocity obstacles with respect to each of the groups but is also closest to the preferred velocity, i.e., 
\begin{align}
\mathbf v_a^{\text{adapt}} = \argmin_{\mathbf v \notin \bigcup_{G \in \{G^i\} - G_a} VO_{a|G}^{\tau}} \|\mathbf v - \mathbf v_a^{\text{pref}}\|,
\end{align}
where $VO_{a|G}^{\tau}$ is the velocity obstacle for agent $a$ induced by the group $G$, as defined in Equation~\ref{eq:VOaG}.

Once the leader's new preferred velocity is computed, we can calculate the  preferred velocity for all other agents in the group $G_a$ iteratively. In particular, we first compute all the agents $\{b\}$ that follow the leader $a$, and the new preferred velocity $\mathbf v_b^{\text{adapt}}$ is set by projecting the old preferred velocity along the direction $\mathbf p_b - \mathbf p_a$.
Once the adapted preferred velocity is computed for each agent in $\{b\}$, we continue to find the new velocity for the followers. This iterative process continues until we compute the new preferred velocities for all agents in the group (see Figure~\ref{fig:meso}).

\subsubsection{Agent-Agent Collision Avoidance}
The new preferred velocity computed is used as the input to the ORCA agent-agent collision avoidance algorithm~\protect\cite{Jur:2011:RVO} that finally computes the current velocity for each agent. The ORCA algorithm ensures the agent avoids collisions with nearby individual agents. The agent need only avoid pairwise collisions with immediately neighboring agents. This computation is performed independently for each agent.

%% file: experiment.tex
\section{Implementation and Performance}
\label{sec:experiment}
In this section we describe our implementation and highlight the performance of our algorithm on different benchmarks.   We compare our result with the grouping behaviors generated by two state-of-the-art crowd simulators:  agent-agent collision avoidance algorithm OCRA~\cite{Jur:2011:RVO} and a group-based {\em  meso-scale} navigation approach~\cite{He:2013:ICRA}. We use five benchmarks to evaluate our algorithms and three of them are designed from real-world videos, and we compare the movement trajectories generated by different approaches; and two other synthetic benchmarks, where we also compare the running time and the number of collisions between the agents during the simulation.  We have implemented our algorithms in C++ on an Intel Core i7 CPU running at 3.30GHz with 16GB of RAM and running Windows 7. All the timing results are generated on a single core. In practice, our 

\subsection{Real-World Scenarios and Validation}
\begin{figure*}
\centering
   \subfloat[real-world video frame]{\includegraphics[width=0.228\linewidth]{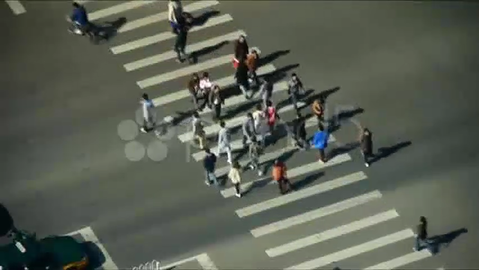}}
   \subfloat[ORCA]{\includegraphics[width=0.24\linewidth]{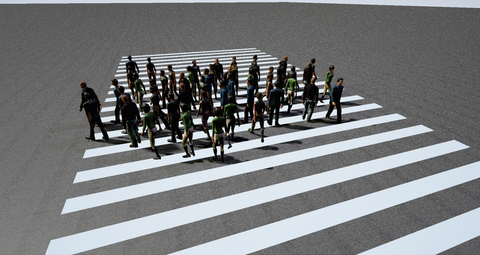}}
   \subfloat[meso-scale]{\includegraphics[width=0.24\linewidth]{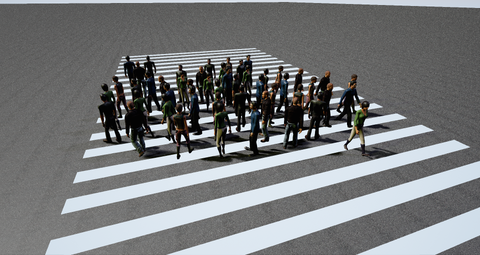}}
   \subfloat[our method]{\includegraphics[width=0.24\linewidth]{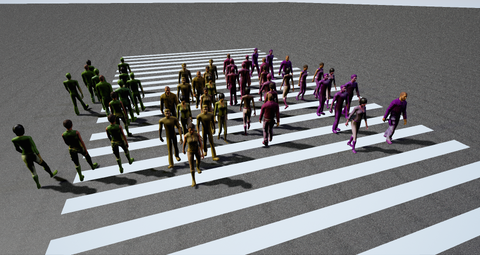}}
   \caption{For benchmark 1, we compare the group behavior generated by our algorithm (d) on a real-world scenario (a). As compared to ORCA (b) and meso-scale (c), our approach can generate smoother and coherent trajectories.}
   \label{fig:benchmark1Comp}
\end{figure*}

\begin{figure*}
\centering
   \subfloat[real-world video frame]{\includegraphics[width=0.228\linewidth]{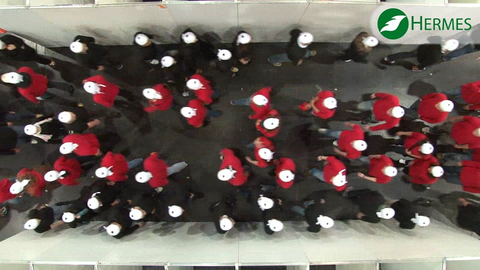}}
   \subfloat[ORCA]{\includegraphics[width=0.24\linewidth]{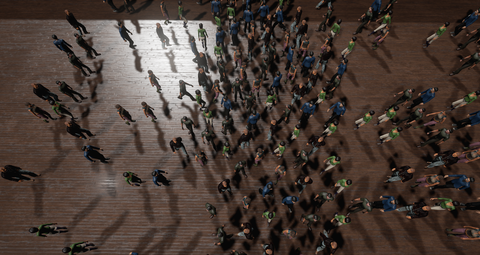}}
   \subfloat[meso-scale]{\includegraphics[width=0.24\linewidth]{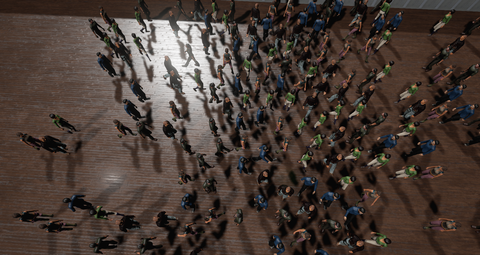}}
   \subfloat[our method]{\includegraphics[width=0.24\linewidth]{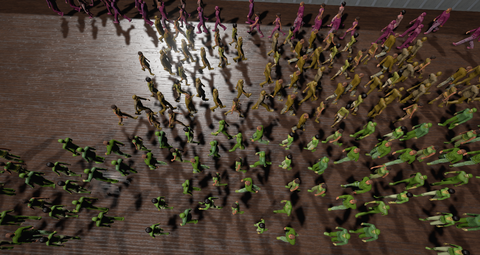}}
   \caption{Comparison between the key frame for simulation sequences generated using different approaches on benchmark 3.}
   \label{fig:benchmark3Comp}
\end{figure*}

\begin{figure*}
\centering
   \subfloat[real-world video frame]{\includegraphics[width=0.21\linewidth]{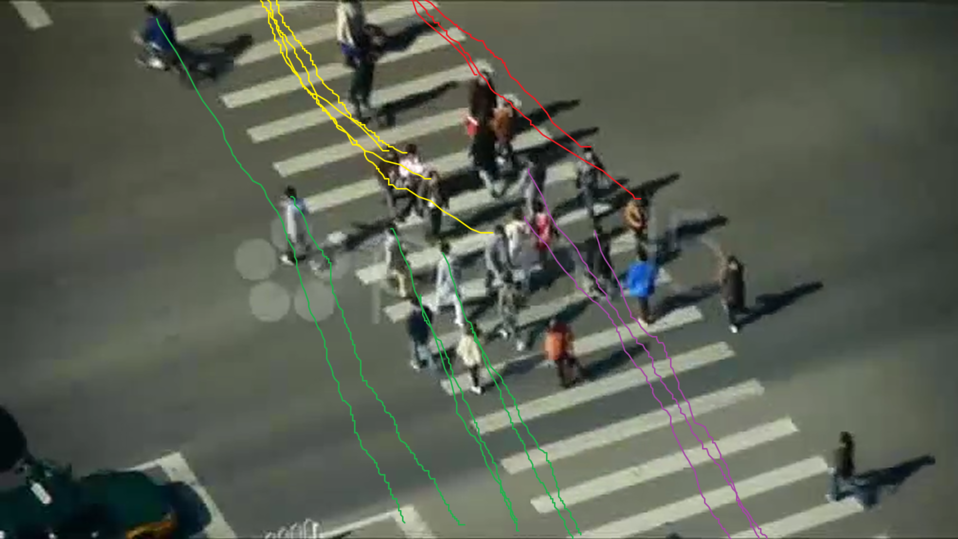}}
   \subfloat[ORCA]{\includegraphics[width=0.24\linewidth]{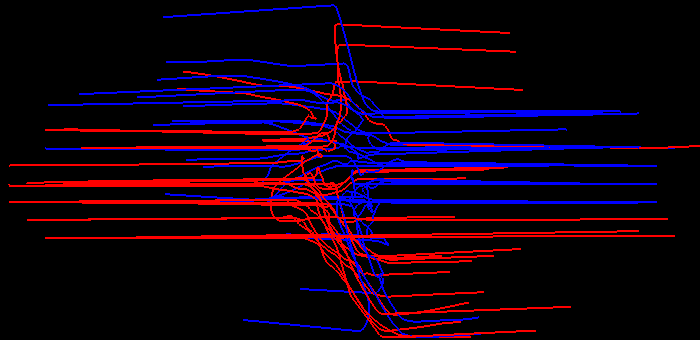}}
   \subfloat[meso-scale]{\includegraphics[width=0.24\linewidth]{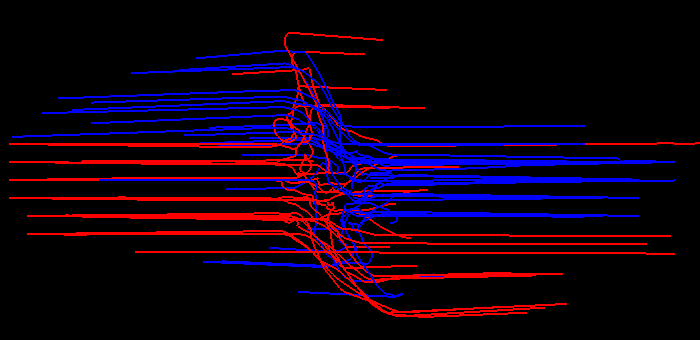}}
   \subfloat[our method]{\includegraphics[width=0.24\linewidth]{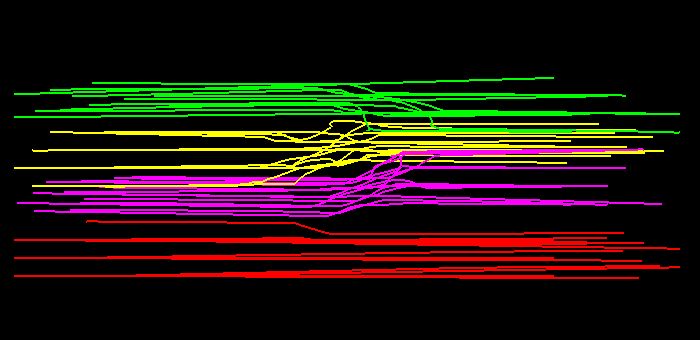}}
   
   \subfloat[real-world video frame]{\includegraphics[width=0.21\linewidth]{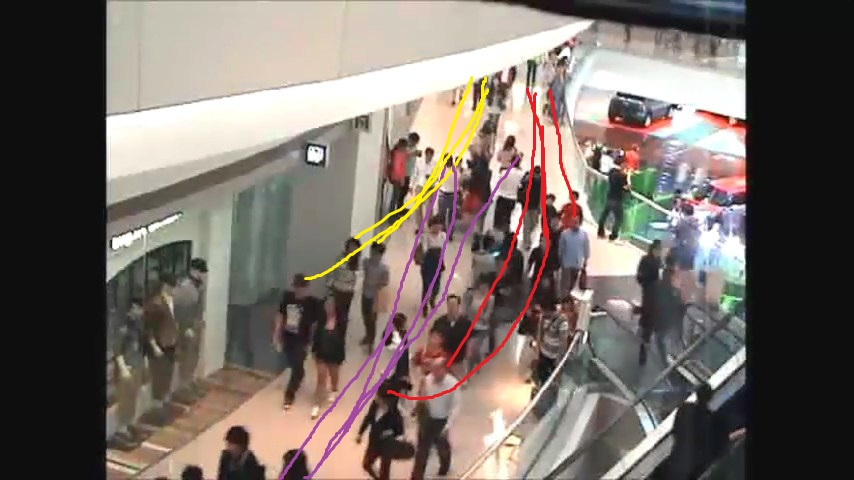}}
   \subfloat[ORCA]{\includegraphics[width=0.24\linewidth]{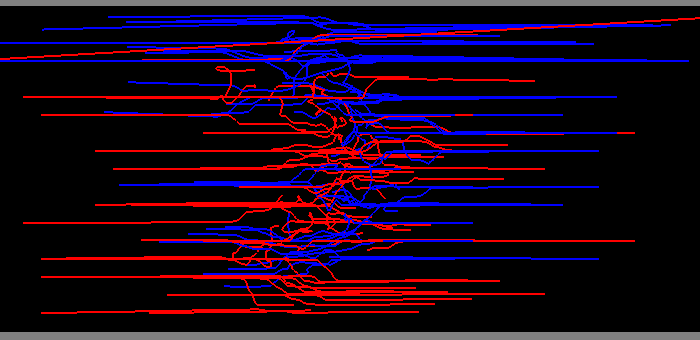}}
   \subfloat[meso-scale]{\includegraphics[width=0.24\linewidth]{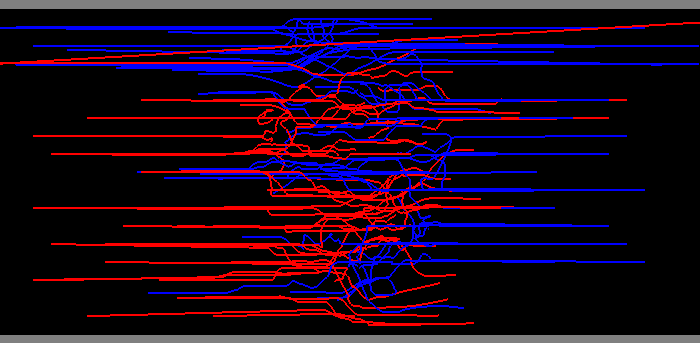}}
   \subfloat[our method]{\includegraphics[width=0.24\linewidth]{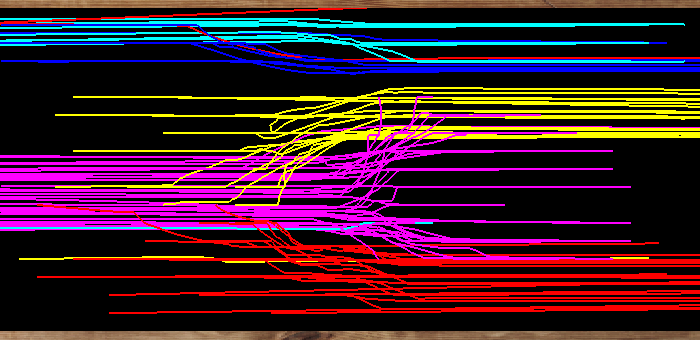}}

   \subfloat[real-world video frame]{\includegraphics[width=0.21\linewidth]{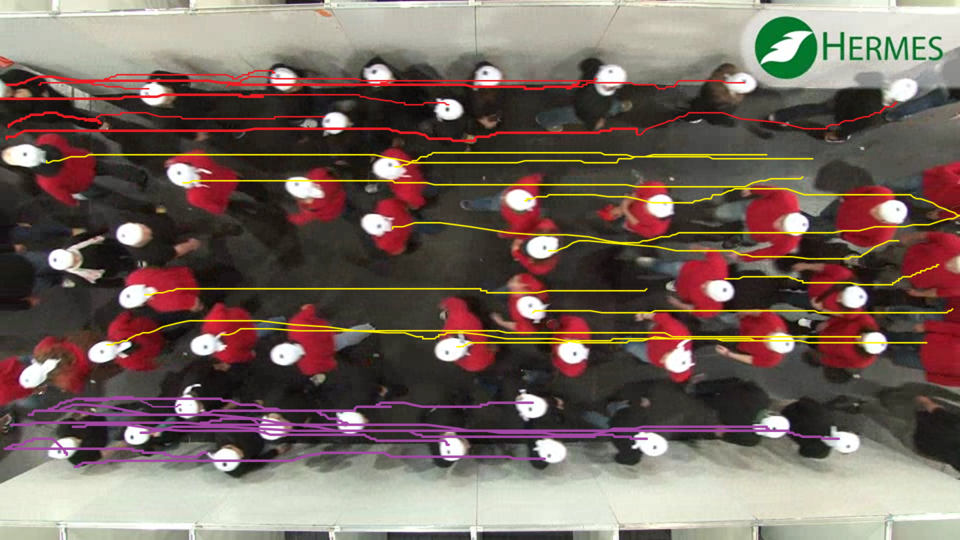}}
   \subfloat[ORCA]{\includegraphics[width=0.24\linewidth]{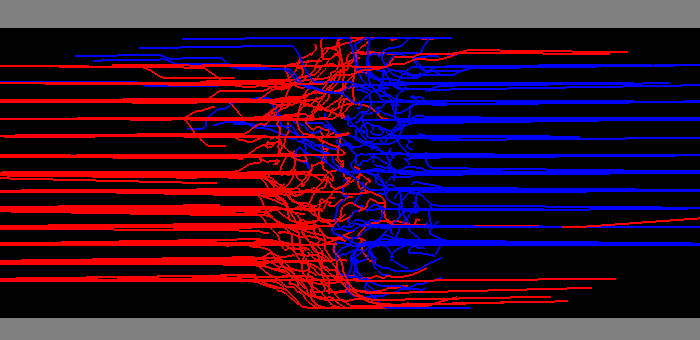}}
   \subfloat[meso-scale]{\includegraphics[width=0.24\linewidth]{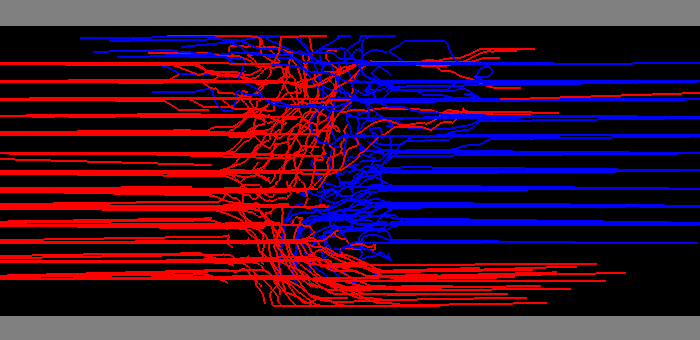}}
   \subfloat[our method]{\includegraphics[width=0.24\linewidth]{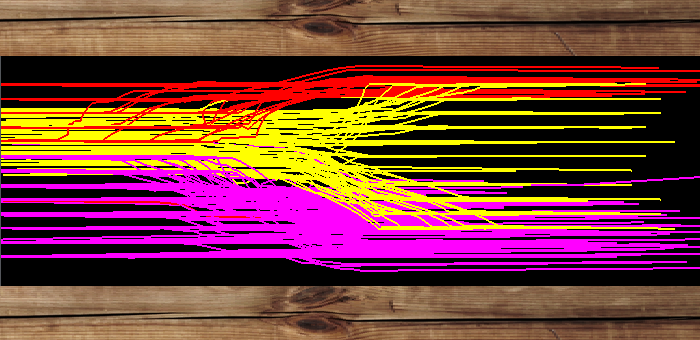}}
   
   \caption{For all the three real-world benchmarks  (the 1st column), we compare the trajectory behaviors generated by our algorithm (4th column, each color represents a group). As compared to ORCA (2nd column) and meso-scale (3rd column), our approach can generate smoother and coherent trajectories.}
   \label{fig:benchmarkCompTraj}
\end{figure*}

We compare the crowd simulation results using our dynamic group behavior generation algorithm and prior approaches on scenarios inspired by real-world crowd videos.  
We extract the trajectories of the agents in the real-world video using a pedestrian tracking algorithm~\cite{Bera:2015:ETE}. For each crowd simulation algorithm, the number of agents and their initial positions and goal positions are assigned according to the pedestrian tracking results. Given the initial and goal positions, we compare the trajectories of the pedestrians generated by each algorithm and compare them with those in the real videos in Figure~\ref{fig:benchmarkCompTraj}. Figures~\ref{fig:benchmark1Comp} and~\ref{fig:benchmark3Comp} show the key frame for simulation sequences generated using different approaches. We can observe that the simulation results using our dynamic group generation algorithm is most similar to the real world pedestrians in terms of trajectory behaviors. 

In terms of  quantitative comparison, we evaluate the behavior of real pedestrians with that of simulated crowds by checking:
\begin{enumerate}
\item Compare the running time and number of collisions that occurred during the navigation from the initial to the goal positions, as shown in Table~\ref{tab:experiment}.
\item Compare the trajectories extracted using the tracking algorithm (i.e. the ground truth) for some of the agents with the trajectories computed by different multi-agent simulation algorithms.
%, as shown in Figure~\ref{fig:videovelocity}. 
\end{enumerate}

\begin{table*}[tbp]
  \centering
  \resizebox{0.8\textwidth}{!}{
    \begin{tabular}{|r|r|r|r|r|r|r|r|r|r|}
    \hline
 \multicolumn{2}{|c|}{Benchmark 1} & \multicolumn{2}{c|}{Benchmark 2} & \multicolumn{2}{c|}{Benchmark 3} & \multicolumn{2}{c|}{Benchmark 4} & \multicolumn{2}{c|}{Benchmark 5}\\
    \hline
    \#agents & \#groups &\#agents & \#groups  & \#agents & \#groups & \#agents & \#groups  & \#agents & \#groups  \\
    \hline
     
      49 & 8 & 97 & 9 & 185 & 6 &184 & 11 & 151 & 9 \\
    \hline
    \end{tabular}
    }
  \caption{Number of agents and maximum number of groups in each benchmark. The number of groups change during the simulation.}
  \label{tab:numofAgents}
\end{table*}

In the first benchmark, agents are passing through a crosswalk as shown in Figure~\ref{fig:benchmark1}. During this simulation, agents automatically aggregate into groups and perform group-level collision avoidance. In this benchmark, the total time taken by different crowd simulation approaches is almost similar.
However, our dynamic group behavior approach result in fewer collisions between the agents during navigation.
Moreover, the trajectories generated using our algorithm have a better match with the ground truth data, as shown in Figure~\ref{fig:benchmark1Comp}. This is due to the fact that ORCA and meso-scale simulation algorithms need more space to perform collision avoidance and therefore the agents are more spread out.

In the second benchmark, agents are moving in a hallway inside the building, which represents a tight space. In this simulation, each agent's initial position and direction of movement is computed based on the real-world trajectories.
 Our approach can compute the navigation trajectories with a few collisions with coherent grouping behaviors, similar to real-world videos.
 In contrast, the agents in  ORCA and  meso-scale simulation algorithms take more time to move from the initial to the goal positions due to the tight spaces.   Moreover, the trajectories computed by our algorithm are smoother and there is high co-relation with the ground truth data, i.e. the extracted trajectories.

\begin{figure*}
\centering
   \includegraphics[width=0.31\linewidth]{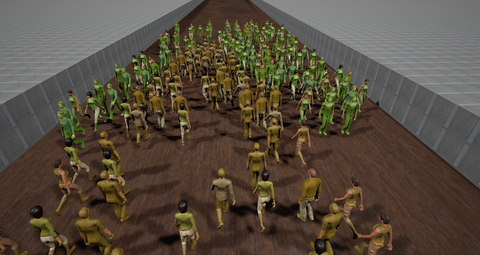}
   \includegraphics[width=0.31\linewidth]{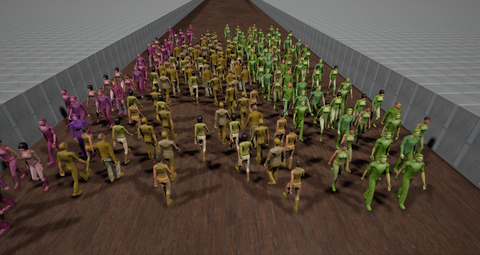}
   \includegraphics[width=0.31\linewidth]{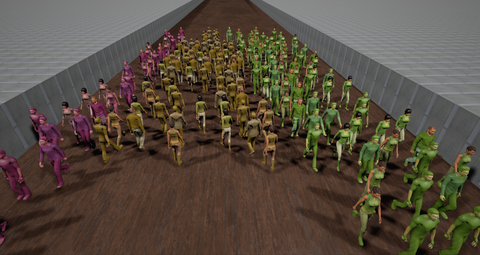}
   \caption{Key frames for the simulation sequence generated by our method on benchmark 3.}
   \label{fig:benchmark3}
\end{figure*}

The third benchmark corresponds to a cluttered environment where the agents need to go through the hallway, as shown in Figure~\ref{fig:benchmark3}. Both RVO and meso-scale methods are not able to compute collision-free navigation as the crowd density is high. Instead, our method automatically enables the agents to move in groups and compute collision-free trajectories. We also observe that the trajectories computed using our algorithm have a better match with the ground truth data.

%\begin{figure*}
%\centering
%   \includegraphics[width=0.31\linewidth]{figs/benchmark/benchmark5-1}
%   \includegraphics[width=0.31\linewidth]{figs/benchmark/benchmark5-2}
%   \includegraphics[width=0.31\linewidth]{figs/benchmark/benchmark5-3}
%   \caption{Key frames for the simulation sequence generated by our method on benchmark 5.}
%   \label{fig:benchmark5}
%\end{figure*}

%\begin{figure}[!ht]
%\centering
%\subcaptionbox{benchmark1}{\includegraphics[width=\linewidth]{figs/Experiment/video1.pdf}}
%\subcaptionbox{benchmark2}{\includegraphics[width=\linewidth]{figs/Experiment/video2.pdf}}
%\subcaptionbox{benchmark3}{\includegraphics[width=\linewidth]{figs/Experiment/video3.pdf}}
%\caption{Speed v.s. density plots for simulated crowds with different approaches on real pedestrian benchmarks. }
%\label{fig:videovelocity}
%\end{figure}

\begin{table*}[tbp]
  \centering
  \resizebox{\textwidth}{!}{
    \begin{tabular}{|c|r|r|r|r|r|r|r|r|r|r|r|r|r|r|r|}
    \hline
    \multicolumn{1}{|c|}{\multirow{2}{*}{Method}} & \multicolumn{3}{c|}{Benchmark 1} & \multicolumn{3}{c|}{Benchmark 2} & \multicolumn{3}{c|}{Benchmark 3} & \multicolumn{3}{c|}{Benchmark 4} & \multicolumn{3}{c|}{Benchmark 5}\\
    \cline{2-16} 
    \multicolumn{1}{|c|}{} & tpf & \#steps & \#colls & tpf & \#steps & \#colls & tpf & \#steps & \#colls & tpf & \#steps & \#colls & tpf & \#steps & \#colls \\
    \hline \hline
     
     ORCA & 3.2 & 162 & 78 & 3.3 & 363 & 103 & 4.2 &  5000+ & 500+ & 8.7 & 209 & 257 & 9.2 & 5000+  & 500+ \\
     Meso-scale & 4.7& 17.1& 53& 4.9& 475 & 111 & 8.9 & 5000+  & 500+  & 9.6 & 212 & 262 & 12.9 &  5000+ & 500+ \\
     Our (dynamic grouping) & 4.8 & 161 & 2 & 4.6 & 221 & 6 & 8.1 & 253 & 3 & 9.4 & 203 & 3 & 10.4 & 387 & 2\\
    \hline
    \end{tabular}
    }
  \caption{The comparison between our approach and previous methods (ORCA, meso-scale) on five benchmarks. We report the average running time per frame (tpf) in milli-second, the average number of simulation time steps taken for each agent to reach the goal position (\#steps) and the average number of pairwise collisions (\#colls). These collisions can occur when the conservative collision avoidance schemes can't compute a feasible solution.  In some case, the agents in the ORCA or meso-scale algorithms get stuck resulting in a high number of collisions. Even after $5000$ simulation they have not reached the goal positions. We observe these behaviors with ORCA and meso-scale algorithms on Benchmark 3 and Benchmark 5.}
  \label{tab:experiment}
\end{table*}

\subsection{Other Benchmarks}
We also generated some synthetic scenes to further evaluate the performance of our dynamic group behavior generation algorithm.
In the fourth benchmark, agents are randomly placed in the scenario. Our approach automatically cluster them into groups and generates coherent trajectories. Furthermore, it results in fewer collisions and smoother trajectories.
The fifth benchmark corresponds to adding several static obstacles in the environment corresponding to the fourth benchmark. Our method can compute the paths to the goal position for each agent. On the other hand, the agents get stuck and pushed away from the goal position within ORCA and meso-scale simulation

%% file: conclusion.tex
\section{Limitations, Conclusions and Future Work}

We present a novel multi-agent navigation algorithm that can automatically generate dynamical grouping behaviors. Our approach is general and makes no assumption about the size or shape of the group, and can dynamically adapt to the environment. Moreover, it results in smooth and coherent navigation behaviors as compared to prior multi-agent reciprocal collision avoidance algorithms.  Furthermore, the agent's tend to avoid congestion based on group's follow-the-leader trajectory computation behavior, which is similar to human behaviors observed in real-world behaviors.  We demonstrate its performance on complex benchmarks with a few hundred agents and show that the trajectories generated by our algorithm are similar to those observed in real-world behaviors and exhibit similar group behaviors. Unlike prior group behavior simulation schemes, our approach is adaptive and can model the dynamic behaviors of the agent in response to the environment. 

Our approach has some limitations. It is currently designed for homogeneous agents and the clustering algorithm only takes into account the position and velocity of each agent. We don't account for agents with varying personalities or how they respond to the environment effects or situations or the psychological component corresponding to the concept of personal space that varies along different cultures or the social norms. Our reciprocal group-group collision avoidance algorithm can be conservative as it is implicitly based on the convex hull or extreme agents.

There are many avenues for future work. In addition to overcoming these limitations, we would like to evaluate its performance in complex scenarios with tens of thousands of agents (e.g. sporting events or religious gatherings). We would like to further validate its performance using other metrics, such as comparing with the collective behaviors and fundamental diagrams of real-world crowds. Finally, we would like to combine with macroscopic techniques to simulate very dense crowds.